\documentclass[journal]{IEEEtran}
\usepackage{cite}
\usepackage{amsmath,amssymb,amsfonts}
\usepackage{algorithmic}
\usepackage{graphicx}
\graphicspath{{figs/}{../}}
\usepackage{algorithm,algorithmic}
\usepackage{hyperref}
\hypersetup{hidelinks}
\usepackage{textcomp}
\usepackage{multirow}
\usepackage{booktabs}
\usepackage{bm}
\usepackage{xcolor}
\newif\ifshowtodo\showtodotrue

\def\BibTeX{{\rm B\kern-.05em{\sc i\kern-.025em b}\kern-.08em
    T\kern-.1667em\lower.7ex\hbox{E}\kern-.125emX}}

\begin{document}

\title{Distortion-Corrected Diffusion MRI Using Rotated-View EPI and Joint Field-Map/Image Estimation with Gaussian Primitives}

\author{Wenqi Huang, Zhitao Li, Nan Wang, Yimeng Lin, Mengze Gao, Yurui Qian, Sevgi Gokce Kafali, Xiaozhi Cao, Kawin Setsompop, Daniel Rueckert and Congyu Liao
\thanks{This work was supported in part by the European Research Council (884622), the National Institutes of Health (R01MH116173, R01HD114719, K99EB035178) and TUM Graduate School
Internationalization Support Grant. Corresponding author: Wenqi Huang (wenqi.huang@tum.de). Daniel Rueckert and Congyu Liao are co-last authors.}
\thanks{Wenqi Huang and Sevgi Gokce Kafali are with the Chair for AI in Healthcare and Medicine, Technical University of Munich (TUM) and TUM University Hospital, and the School of Computation, Information and Technology, TUM, Munich, Germany.}
\thanks{Congyu Liao and Yurui Qian are with the Neuroimaging Technology Research Center, Department of Radiology and Biomedical Imaging, University of California, San Francisco, CA, USA.}
\thanks{Zhitao Li is with the Innovation Academy for Precision Measurement Science and Technology, Chinese Academy of Sciences, Wuhan, China.}
\thanks{Nan Wang, Yimeng Lin, Mengze Gao, Xiaozhi Cao, and Kawin Setsompop are with the Department of Radiology, Stanford University, Stanford, CA, USA.}
\thanks{Daniel Rueckert is with the Chair for AI in Healthcare and Medicine, TUM and TUM University Hospital, Munich, Germany, the Munich Center for Machine Learning (MCML), and the Department of Computing, Imperial College London, U.K.}
}

\maketitle
\bstctlcite{IEEEtran:BSTcontrol}

\begin{abstract}
Echo planar imaging (EPI) is the standard acquisition technique for diffusion and functional neuroimaging, enabling rapid imaging but suffering from geometric distortions caused by $B_0$ field inhomogeneities. Existing correction methods first reconstruct distorted images using parallel imaging, then estimate the $B_0$ field and correct the distortion in the image domain. In this sequential process, reconstruction artifacts at high acceleration factors and low SNR at high diffusion $b$-values degrade $B_0$ estimation and limit the overall correction quality. We propose a physics-informed framework that jointly estimates the $B_0$ field, the distortion-free image, and the shot-to-shot phase directly from \textit{k}-space data, without depending on an intermediate parallel-imaging reconstruction. All three are represented as superpositions of Gaussian primitives embedded within an EPI forward model; the explicit, continuous parameterization captures both smooth regions and tissue boundaries and is evaluated on rotated sampling grids without resampling. The diffusion-weighted image is modeled as real and non-negative, with the image phase absorbed into a per-shot phase factor. Acquiring the blip-up/blip-down data at several phase-encoding orientations, the rotated-view principle of PROPELLER applied to full-FOV single-shot EPI, distributes distortions across orientations, improving point spread function isotropy and providing stronger constraints for $B_0$ estimation. On \textit{in vivo} brain diffusion EPI at $R{=}4$ and $5$, the proposed method attains the closest brain-boundary agreement with a distortion-free structural reference, with the largest improvement over sequential methods at high $b$-value.
\end{abstract}

\begin{IEEEkeywords}
Diffusion MRI, Echo Planar Imaging, $B_0$ Inhomogeneity, Gaussian Primitives, Gaussian Splatting, Distortion Correction, Joint Reconstruction
\end{IEEEkeywords}

\section{Introduction}
\label{sec:introduction}
\IEEEPARstart{E}{cho} Planar Imaging (EPI) is a fast MRI acquisition technique widely used for functional and diffusion imaging~\cite{setsompop2012improving} due to its ability to capture an entire image or \textit{k}-space plane within a single or a few excitations. However, the long echo train and correspondingly low effective bandwidth along the phase-encoding direction make EPI highly sensitive to $B_0$ field inhomogeneities, which introduce geometric distortions along that direction. These distortions manifest as stretching, compression, intensity pileup, and spatial misalignment, and are most severe near air-tissue interfaces such as the frontal sinuses and temporal lobes~\cite{andersson2003correct,smith2004advances}. Correcting these artifacts is critical for preserving spatial accuracy in quantitative analysis, tractography, and clinical diagnosis.

A range of methods have been developed to correct $B_0$-induced EPI distortions, and most of them operate in the image domain after the EPI images have been reconstructed. Pre-acquired field maps~\cite{jezzard1995correction} estimate the $B_0$ distribution from a separate scan, but this needs extra acquisition time and can be invalidated by inter-scan motion or by physiological $B_0$ fluctuations from breathing and cardiac pulsation. To avoid the extra scan, reversed phase-encoding (blip-up/blip-down) methods acquire two EPI images with opposite phase-encoding polarities, so that $B_0$-induced distortions appear in opposite directions. TOPUP~\cite{andersson2003correct,smith2004advances} takes the coil-combined blip-up/blip-down image pair and estimates the off-resonance field in the image domain, modeling its effect as a phase-encoding voxel displacement with intensity pileup, and has become the standard correction tool for diffusion MRI in the Human Connectome Project and similar large-scale studies~\cite{sotiropoulos2013advances}. Extensions include EPIC~\cite{holland2010efficient}, DR-BUDDI~\cite{irfanoglu2015drbuddi}, and interlaced q-space schemes~\cite{bhushan2014improved}. Multi-shot acquisitions reduce the echo train length and thus the distortion per shot, with dedicated reconstructions such as MUSSELS~\cite{mani2017multi} resolving the inter-shot phase, but they still require a separate $B_0$ correction step. All of these methods share a common pipeline: first reconstruct the distorted EPI images using parallel imaging (e.g., CG-SENSE~\cite{pruessmann2001advances}) or structured low-rank reconstruction, then estimate the field and correct the distortion in the image domain~\cite{andersson2003correct,andersson2016integrated}.

In this sequential pipeline, $B_0$ estimation is directly limited by the quality of the reconstructed distorted images. At higher acceleration factors, parallel-imaging aliasing and noise amplification grow, propagate into the $B_0$ estimate, and cap how far each EPI shot can be accelerated. Model-based reconstruction folds $B_0$ off-resonance into an iterative forward model~\cite{sutton2003fast}, directly producing a corrected image, but assumes a known or separately measured field map. Conjugate-phase and time-segmentation methods~\cite{noll1991homogeneity,schomberg1999off} and multifrequency interpolation~\cite{man1997multifrequency} approximate the off-resonance term by a few frequency-independent components so that fast transforms apply, and iterative model-based reconstructions~\cite{harshbarger1999iterative,yarach2017model} are accelerated by min-max NUFFT interpolation and Toeplitz embedding~\cite{fessler2005toeplitz}. BUDA (Blip-Up/Down Acquisition)~\cite{liao2019highly,liao2021distortion} combines the blip-up and blip-down multi-coil \textit{k}-space data in a single Hankel-structured low-rank reconstruction, but likewise takes its $B_0$ from a separate TOPUP estimate that is held fixed, and CORPUSE~\cite{xiong2026corpuse} applies the same fixed-field principle to the rotated blades of PROPELLER-EPI. Across these methods, the field map is measured or estimated separately and never jointly optimized with the image, so field-estimation and reconstruction errors compound.

Estimating the field map together with the image from the same \textit{k}-space data removes the separate estimation step. Sutton \textit{et al.}~\cite{sutton2004dynamic} estimated a dynamic field map jointly with the image from spiral-in/spiral-out data, and Matakos and Fessler extended joint estimation to parallel single-shot acquisitions~\cite{matakos2010joint} and to through-plane field gradients~\cite{matakos2011dynamic}. Balachandrasekaran \textit{et al.}~\cite{balachandrasekaran2019calibration} formulated calibration-free EPI field correction as structured low-rank matrix recovery. Building on this formulation, the present work adds a per-shot phase model, an explicit continuous representation, and rotated views.

A different route avoids distortion at acquisition. Echo planar time-resolved imaging (EPTI)~\cite{wang2019epti} samples a spatiotemporal \textit{k}-$t$ space across the echo train and reconstructs a series of distortion- and blurring-free images at closely spaced echo times, resolving the field evolution within the readout instead of correcting for it afterwards. ACE-EPTI~\cite{dong2022acepti} improves its SNR efficiency for diffusion imaging. These methods remove the need for distortion correction at the cost of a dedicated multi-shot sequence and subspace reconstruction. PROPELLER~\cite{pipe1999motion} rotates a blade of parallel phase-encoding lines across shots, uses the oversampled \textit{k}-space center of each blade for motion estimation, and combines the blades by gridding. PROPELLER-EPI~\cite{wang2005propellerepi} acquires each blade with a short EPI readout along the narrow blade axis, which reduces distortion in diffusion imaging. Diffusion-PEPTIDE~\cite{fair2021diffpeptide} rotates the \textit{k}-space encoding of EPTI across shots, as in PROPELLER, and obtains distortion- and blurring-free diffusion images with self-navigated motion correction. ROMER-EPTI~\cite{dong2025romer} and ROVER-dMRI~\cite{liu2026roverdmri} instead rotate the slice orientation of thick-slice volumes and combine them in a super-resolution reconstruction. In these methods the rotation serves motion robustness, SNR efficiency, or resolution, and off-resonance is handled by short blades or by echo-train encoding.

Deep learning has also been applied to EPI distortion correction. Unsupervised methods such as Duong \textit{et al.}~\cite{duong2020unsupervised} and FD-Net~\cite{alkilani2024fdnet} learn the distortion from reversed phase-encoding pairs without ground truth and achieve fast inference. Learning-based correction, however, is tied to its training distribution. It needs protocol-matched data, degrades under domain shift in resolution, FOV, acceleration, or phase-encoding orientation, and as a learned image prior can hallucinate structure unsupported by the measured data~\cite{bhadra2021hallucinations}. Such methods also operate in the image domain on an initial parallel-imaging reconstruction, and thus inherit its acceleration ceiling.

These limitations motivate a scan-specific, physics-informed approach in which we jointly estimate the $B_0$ field and the distortion-free image directly from \textit{k}-space through a single forward model that embeds the full EPI physics, requiring no training data and avoiding the sequential pipeline's error accumulation and its dependence on an intermediate parallel-imaging reconstruction. The estimator already corrects distortion accurately from a single blip-up/blip-down orientation. Acquiring the data at several rotated phase-encoding orientations, the rotated-blade principle of PROPELLER~\cite{pipe1999motion} applied to full-FOV single-shot EPI readouts, further improves point spread function (PSF) isotropy and \textit{k}-space coverage and, because the distortion direction rotates with the phase-encoding axis, conditions the field estimate. We introduced this estimator in a conference version of this work~\cite{huang2025pinr} with a hash-grid implicit neural representation (INR)~\cite{muller2022instant}, but only on synthetic data and without a per-shot phase model. Relative to that version, the present work moves to prospectively acquired \textit{in vivo} data, adds a per-shot phase model, replaces the hash-grid INR with explicit Gaussian primitives, and adds the rotated-view acquisition. The move to \textit{in vivo} data raises two challenges, the spatial representation and the shot-to-shot phase.

Realizing this requires a representation suited to the rotated, off-grid sampling of rotated-view EPI. Discrete voxel grids require interpolation there, which degrades accuracy, while INRs~\cite{mildenhall2021nerf,shen2022nerp,huang2023neural,spieker2023iconik}, including the hash-grid INR of our conference version, store the signal in opaque network weights. We instead use explicit Gaussian primitives~\cite{kerbl20233dgs,peng20253dgsmr,terpstra2026gaussian,singh2026gaussian}, a continuous, interpolation-free representation whose interpretable, directly controllable parameters capture both smooth fields and sharp boundaries, for the image, the smooth $B_0$ field, and the per-shot phase factor introduced next.

\textit{In vivo} data add a shot-to-shot phase absent from the synthetic conference setting. Physiological noise, mainly cardiac pulsation and respiration, induces a different, spatially varying phase on each excitation, largest at high $b$. We model the diffusion-weighted image as real and non-negative, and absorb this phase into a per-shot phase factor, which avoids the signal bias of magnitude reconstruction at low SNR.

Our key contributions are as follows:
\begin{enumerate}
\item A joint \textit{k}-space estimation framework for rotated-view EPI. Rotating the phase-encoding direction across shots originates with PROPELLER~\cite{pipe1999motion}; building on it, we acquire complete, full-FOV single-shot EPI readouts at several orientations and pass all of them, without per-view reconstruction or image-domain combination, to one estimator in which the rotating distortion direction supplies complementary constraints on $B_0$. The rotated views also improve the effective point spread function.
\item A physics-informed framework that jointly estimates the $B_0$ field and the distortion-free image directly from \textit{k}-space, removing the sequential pipeline's dependence on an intermediate parallel-imaging reconstruction. The image, the $B_0$ field, and the per-shot phase factor all share a single explicit, continuous Gaussian-primitive backbone that handles rotated-view grids without interpolation and exposes interpretable, directly controllable parameters.
\item An \textit{in vivo} signal model for this framework: a real, non-negative image with an explicit per-shot phase factor that handles the shot-to-shot physiological phase without magnitude bias, together with a $B_0$ smoothness prior that, with multiple orientations, reduces the single-orientation $B_0$/per-shot-phase ambiguity at high $b$.
\item Validation on two \textit{in vivo} brain diffusion data sets. On the first, across orientations, acceleration factors ($R{=}4$, $5$), and $b$-values, the proposed method attains the closest brain-boundary agreement, with the largest gains over sequential methods at high $b$-value ($b{=}2500$\,s/mm$^2$); at matched shot count, rotated blip-up-only views constrain the field as well as paired views. On the second, a whole-brain 30-direction acquisition at $R{=}3$, we compare DTI parameter maps, their precision, and CSD fiber orientations with their bootstrap uncertainty against TOPUP, BUDA, and the hash-grid INR.
\end{enumerate}

\section{Method}
\label{sec:method}

\begin{figure*}[!tbp]
    \centering
    \includegraphics[width=\linewidth]{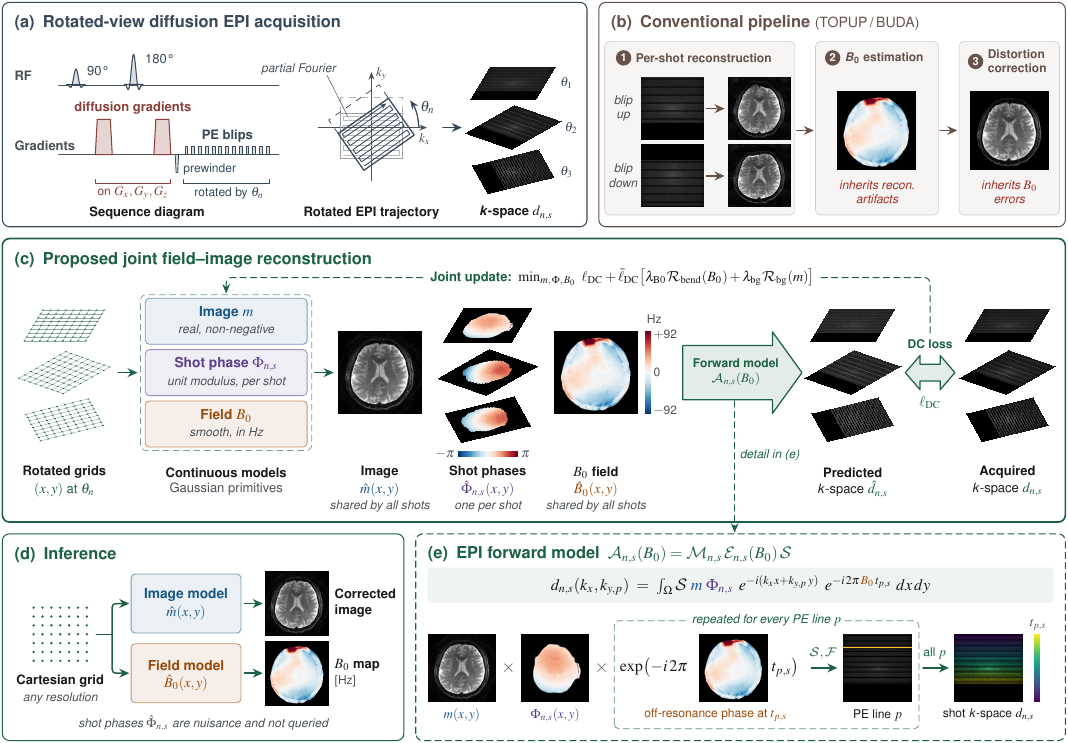}
    \caption{Overview of the proposed method.
        \textbf{(a)}~Rotated-view diffusion EPI: each shot is a single-shot EPI readout whose in-plane encoding axes are rotated by $\theta_n$, so multiple phase-encoding orientations give complementary distortion constraints.
        \textbf{(b)}~Conventional pipeline (TOPUP/BUDA): per-shot reconstruction, $B_0$ estimation, then distortion correction, so reconstruction artifacts propagate into the field estimate and field errors into the corrected image.
        \textbf{(c)}~Proposed joint reconstruction: the image $\hat{m}$ and field $\hat{B}_0$ (shared by all shots) and the per-shot phases $\hat{\Phi}_{n,s}$ are continuous Gaussian-primitive models, sampled on the rotated grids and passed through the EPI forward model $\mathcal{A}_{n,s}(B_0)$; all three are updated jointly by minimizing the \textit{k}-space data-consistency loss $\ell_{\mathrm{DC}}$ with the regularizers of~\eqref{eq:loss}.
        \textbf{(d)}~Inference: the trained image and field models are queried on a Cartesian grid of any resolution.
        \textbf{(e)}~EPI forward model: each phase-encoding line $p$ of shot $(n,s)$ is the Fourier encoding of the coil-weighted magnetization $\mathcal{S}\,m\,\Phi_{n,s}$ modulated by the off-resonance phase accrued at its echo time $t_{p,s}$.}
    \label{fig:method_overview}
\end{figure*}

\subsection{Problem Formulation}
\label{sec:formulation}

EPI samples \textit{k}-space along a zig-zag trajectory driven by alternating readout gradients. With a homogeneous $B_0$ field, the acquired data yield a distortion-free image by inverse Fourier transform; off-resonance from $B_0$ inhomogeneity introduces a time-dependent phase that distorts the reconstruction along the phase-encoding direction. The standard countermeasure is a blip-up/blip-down acquisition, two single-shot EPI readouts of opposite phase-encoding polarity, whose $B_0$-induced distortions point in opposite directions and together constrain the field. Each readout is a separate RF excitation, and inter-shot motion, from physiological noise and bulk head motion, imprints a spatially varying phase that differs from shot to shot. We index each shot by an orientation $n$ and a shot/polarity $s$ (blip polarity $\rho_s\in\{+,-\}$). The shot-to-shot motion imparts a per-shot phase $\varphi_{n,s}(x,y)$, so the magnetization seen by shot $(n,s)$ is $m(x,y)\,\Phi_{n,s}(x,y)$ with $\Phi_{n,s} = e^{i\varphi_{n,s}}$, where $m(x,y)$ is the shared distortion-free image. The acquired multi-coil \textit{k}-space data is
\begin{equation}
\begin{split}
d_{n,s}(k_x, k_{y,p}) = \int_{\Omega} & \mathcal{S}(x,y)\, m(x, y)\, \Phi_{n,s}(x,y) \\
  & \cdot e^{-i (k_x x + k_{y,p} y)}\, e^{-i\, 2\pi\, B_0(x,y)\, t_{p,s}}\, dx\, dy,
\end{split}
\end{equation}
where \(\Omega\) is the image domain, \(\mathcal{S}(x,y)\) is the coil sensitivity, a known quantity that can be obtained from a low-resolution gradient-echo (GRE) pre-scan, and \(B_0(x,y)\) is the off-resonance frequency in Hz, \(B_0 = \gamma\,\Delta B(x,y)/(2\pi)\) with \(\gamma\) the gyromagnetic ratio in rad\,s\(^{-1}\)\,T\(^{-1}\). Phase-encoding line \(p\) of shot \(s\) is sampled at echo time \(t_{p,s}\), which advances by the echo spacing with sign set by the blip polarity $\rho_s$, so $e^{-i\,2\pi B_0(x,y)\,t_{p,s}}$ is the accrued off-resonance phase. We write this forward model compactly as $d_{n,s} = \mathcal{A}_{n,s}(B_0)\,(m\,\Phi_{n,s})$, where the per-shot operator factorizes as $\mathcal{A}_{n,s}(B_0) = \mathcal{M}_{n,s}\,\mathcal{E}_{n,s}(B_0)\,\mathcal{S}$: $\mathcal{S}$ expands the image into coil channels, $\mathcal{M}_{n,s}$ selects the acquired phase-encoding (PE) lines, and $\mathcal{E}_{n,s}(B_0)$ is the off-resonance Fourier encoding that maps a coil image $g$ to
\begin{equation}
\begin{split}
[\mathcal{E}_{n,s}(B_0)\,g](k_x,k_{y,p}) ={}& \int_\Omega g(x,y)\,e^{-i(k_x x + k_{y,p} y)} \\
&\cdot\, e^{-i\,2\pi B_0(x,y)\,t_{p,s}}\,dx\,dy.
\end{split}
\label{eq:encoding}
\end{equation} The $B_0$-induced phase produces geometric distortion along the phase-encoding direction; the per-shot phase, if uncorrected, causes signal cancellation and artifacts when the shots are combined.

The image $m$ is the coil-combined object. For each of the $N_c$ receive coils, $\mathcal{S}$ multiplies the shot magnetization $m\,\Phi_{n,s}$ by that coil's sensitivity $\mathcal{S}_c(\mathbf{r})$ before the encoding $\mathcal{E}_{n,s}$ and the line selection $\mathcal{M}_{n,s}$, and the data term below sums the residual over all coils. The sensitivities are estimated once with ESPIRiT~\cite{uecker2014espirit} from a GRE calibration scan and held fixed. They are neither represented by primitives nor estimated jointly.

The blip-up/blip-down pair uses a single phase-encoding orientation. Rotated-view EPI generalizes the acquisition to $N_{\mathrm{rot}}$ orientations $\theta_1,\ldots,\theta_{N_{\mathrm{rot}}}$, acquiring at each either a blip-up/blip-down pair or a single blip-up readout; writing $\mathcal{P}=\{+,-\}$ for a pair and $\mathcal{P}=\{+\}$ for blip-up only, a reconstruction uses $N_{\mathrm{shot}}=N_{\mathrm{rot}}|\mathcal{P}|$ single-shot excitations, with the standard pair recovered as $N_{\mathrm{rot}}=1$, $\mathcal{P}=\{+,-\}$. Each view is a full-FOV single-shot EPI readout acquired on its own Cartesian grid in the rotated logical gradient frame: sampling within a view is Cartesian, and only the relative orientation between views is not. Distributing the $B_0$-induced distortions across orientations strengthens the constraints on $B_0$; the multi-orientation \textit{k}-space coverage also improves PSF isotropy and suppresses noise by lowering the noise amplification of the parallel-imaging encoding (brain-mean g-factor $2.18$ for one orientation at $R{=}5$, $1.43$ for two and $1.07$ for five, whereas repeating one orientation leaves it unchanged) (Fig.~\ref{fig:psf_analysis}). The joint $B_0$, image, and shot-phase estimation problem across all shots is:
\begin{equation}
\begin{split}
\hat{B}_0, \hat{m}, \hat{\Phi} = \arg\min_{B_0,m,\Phi}
  & \sum_{n,s} \big\| d_{n,s} - \mathcal{A}_{n,s}(B_0)\,(m\,\Phi_{n,s}) \big\|_2^2 \\
  & + \mathcal{R}(B_0),
\end{split}
\label{eq:jointopt}
\end{equation}
where $\mathcal{R}(B_0)$ regularizes the field via its bending energy~\eqref{eq:bending} and $\Phi \triangleq \{\Phi_{n,s}\}_{n,s}$ collects the per-shot phases. The difference from conventional sequential methods is that $B_0$, $m$, and $\Phi$ are optimized jointly through the physics forward model (Fig.~\ref{fig:method_overview}).

With a single phase-encoding orientation the inverse problem is ill-conditioned because the $B_0$-induced distortion acts along one axis only, where a smooth $B_0$ perturbation can be partly absorbed into the image or the per-shot phase with little change to the predicted \textit{k}-space, most acutely at high $b$ (low SNR, large per-shot phase). Additional orientations condition the problem, since each aligns the distortion with a different axis, so a perturbation absorbed under one orientation leaves a visible residual under another.

\subsection{Gaussian Primitive Representations}
\label{sec:primitives}

Solving the joint optimization in~\eqref{eq:jointopt} requires choosing a representation for the image \(m(\mathbf{r})\) and $B_0$ field \(B_0(\mathbf{r})\), with \(\mathbf{r}=(x,y)\). Because rotated-view acquisitions sample coordinates on rotated grids, a continuous representation is preferred so that the image and $B_0$ field can be evaluated directly at the rotated coordinates rather than reconstructed on a fixed grid and resampled, which would introduce interpolation error. Since sampling within a view is Cartesian, the time-segmented, multifrequency, and NUFFT-based methods discussed in Section~\ref{sec:introduction} would correct each view directly if $B_0$ were known; here $B_0$ is itself being estimated, and the resampling error enters the gradient with respect to $B_0$ at every iteration. We represent all three model fields with explicit Gaussian primitives: the image and the $B_0$ field (below) and the per-shot phase (Section~\ref{sec:phasor}).

A 2D Gaussian primitive centered at \(\bm{\mu} \in \mathbb{R}^2\) is
\begin{equation}
G(\mathbf{r}) = \exp\!\Big(
  -\tfrac{1}{2}(\mathbf{r}-\bm{\mu})^\top
  \bm{\Sigma}^{-1}
  (\mathbf{r}-\bm{\mu})\Big),
\label{eq:gaussian}
\end{equation}
where the covariance matrix \(\bm{\Sigma} = \mathbf{R}(\alpha)\,\mathrm{diag}(\sigma_{1}^2, \sigma_{2}^2)\,\mathbf{R}(\alpha)^\top\) is parameterized by a rotation angle \(\alpha\) and anisotropic scales \((\sigma_{1}, \sigma_{2})\). Each primitive thus has five shape parameters, namely position \(\bm{\mu}\) (2), log-scales \(\log \sigma_{k}\) (2), and orientation \(\alpha\) (1), together with a weight whose type depends on the field it represents (below).

The diffusion-encoded image is complex, \(m(\mathbf{r})\,e^{i\varphi_{n,s}(\mathbf{r})}\), with a phase \(\varphi_{n,s}\) dominated by a spatially smooth, shot-to-shot background from physiological noise that carries no diffusion contrast~\cite{eichner2015real}. Conventional pipelines discard it by taking the magnitude, but at the low SNR of high-$b$, high-resolution acquisitions the magnitude of a low-signal voxel in zero-mean complex Gaussian noise follows a Rician distribution (Rayleigh in the no-signal limit) and develops a positive floor that biases the signal upward. We instead never form a magnitude; the measured \textit{k}-space is kept complex and the smooth per-shot background phase is absorbed into $\Phi_{n,s}$ (Section~\ref{sec:phasor}), so the data term is a complex least-squares fit whose residual stays zero-mean Gaussian and floor-free. We nonetheless model \(m\) as real and non-negative as a prior on the noise-free signal~\cite{eichner2015real}, enforced through the parameterization. With non-negative weights \(w_j \ge 0\) and a strictly positive Gaussian kernel, \(m = \sum_j w_j G_j\) in~\eqref{eq:models} satisfies \(m \ge 0\) at every query resolution. This also reduces one branch of the \(B_0\)\,/\,image-phase\,/\,per-shot-phase identifiability degeneracy of Section~\ref{sec:formulation}, since a global image sign flip can no longer be reabsorbed into the per-shot phases. The \(B_0\) field uses signed real weights, and $\Phi_{n,s}$ is represented by the two real fields described below.

We represent the image as a superposition of \(N_{\mathrm{img}}\) Gaussian primitives and the $B_0$ field as a superposition of \(N_{\mathrm{B0}}\) primitives, the $j$-th ($G_j$) carrying its own center, scales, and orientation:
\begin{equation}
m(\mathbf{r}) = \sum_{j=1}^{N_{\mathrm{img}}} w_j\, G_j(\mathbf{r}), \quad
B_0(\mathbf{r}) = \sum_{j=1}^{N_{\mathrm{B0}}} w_j^{(\mathrm{B0})}\, G_j(\mathbf{r}),
\label{eq:models}
\end{equation}
with image weights \(w_j \ge 0\). Since both representations are continuous over \(\mathbb{R}^2\), rotated-view acquisitions are handled by evaluating the primitives on rotated coordinate grids \(\mathbf{r}' = \mathbf{R}(\theta_n)\,\mathbf{r}\).

\subsection{Per-Shot Phase Parameterization}
\label{sec:phasor}

The per-shot phase is most pronounced at high diffusion $b$-values, where cardiac pulsation about the ventricles~\cite{miller2003nonlinear} imparts a localized, spatially nonlinear phase that can exceed $\pm\pi$, and eddy currents from the strong diffusion gradients add further shot-dependent phase. A common approach estimates a smooth scalar phase field $\varphi_{n,s}(\mathbf{r})$ (from a navigator or low-resolution phase image, or fit to a low-order smooth basis) and applies the phase factor $\Phi_{n,s} = e^{i\varphi_{n,s}}$. When the phase exceeds $\pm\pi$, however, phase wrapping introduces $2\pi$ discontinuities that a smooth scalar field cannot represent; these wrap boundaries are numerically unstable and degrade the reconstruction.

We instead parameterize the phase factor directly as a unit-modulus complex field
\begin{equation}
\Phi_{n,s}(\mathbf{r}) = \frac{u_{n,s}(\mathbf{r}) + i\, v_{n,s}(\mathbf{r})}{\sqrt{u_{n,s}(\mathbf{r})^2 + v_{n,s}(\mathbf{r})^2}\,},
\label{eq:phasor}
\end{equation}
where $u_{n,s}$ and $v_{n,s}$ are the real and imaginary parts of a basis of $N_\varphi$ complex-weighted Gaussian primitives owned by shot $(n,s)$. Each shot has its own basis, and within a shot the two fields share the primitive positions, scales, and orientations, differing only in their real and imaginary weights. In implementation, the denominator is clamped by a small $\epsilon$ for numerical stability. Because $\Phi_{n,s}$ is a continuous unit-modulus field rather than a wrapped scalar, it is free of phase wrapping, since smooth motion phase exceeding $\pm\pi$ is represented by a winding of $(u,v)$ with no discontinuity. The Gaussian basis provides compact local support that captures the spatially localized phase from cardiac pulsation and eddy currents more efficiently than smooth global polynomials.

\subsection{Optimization and Inference}
\label{sec:optimization}

Substituting the primitive models~\eqref{eq:models} into~\eqref{eq:jointopt}, the unknowns become the Gaussian-primitive parameters of all three fields (the position, anisotropic scales, orientation, and weight of every image, $B_0$, and per-shot phase primitive), which we fit jointly by minimizing
\begin{equation}
\begin{split}
\mathcal{L} & = \sum_{n,s} \big\| d_{n,s} - \mathcal{A}_{n,s}(B_0)\,(m\,\Phi_{n,s}) \big\|_2^2 \\
  & \quad + \lambda_{\mathrm{B0}}\,\mathcal{R}_{\mathrm{bend}}(B_0).
\end{split}
\label{eq:loss}
\end{equation}
Optimizing these parameters updates the image, $B_0$, and per-shot phase together. The data-consistency term is the squared-$\ell_2$ \textit{k}-space residual, the maximum-likelihood objective for approximately white, zero-mean complex-Gaussian acquisition noise. A small image penalty is added to~\eqref{eq:loss} and, like the bending energy, scaled by the current data-consistency residual: a background penalty, the squared image magnitude outside the head support of the GRE calibration image, that suppresses aliasing leaking into the air ($\lambda_{\mathrm{bg}}$ in Fig.~\ref{fig:method_overview}c).

\noindent\textbf{Initialization and scale floor.}
The $B_0$ field and per-shot phase are far smoother than the image and use correspondingly fewer primitives. Image primitive positions $\bm{\mu}_j$ are sampled uniformly over the head support (the smoothed GRE magnitude thresholded into a binary mask), with small random initial weights. The $B_0$ and per-shot phase primitives are seeded the same way, but their small count makes a Poisson-disk minimum-distance constraint~\cite{bridson2007fast} affordable, which prevents sub-resolution primitive collisions and keeps these smooth fields evenly covered. A fixed minimum scale floor $\sigma_{\min}$ on the image primitives caps the per-primitive frequency content and prevents overfitting to noise.

\noindent\textbf{Bending-energy regularization on $B_0$.}
Single-orientation acquisitions retain the ambiguity of Section~\ref{sec:formulation}. Without an external prior, the optimizer can split residual phase between $B_0$ and the per-shot phase factor, and $B_0$ can drift to a fragmented field (the real image carries no phase of its own, Section~\ref{sec:primitives}). We constrain $B_0$ by its bending energy~\cite{rueckert1999nonrigid}:
\begin{equation}
\mathcal{R}_{\mathrm{bend}}(B_0) = \int_\Omega (\partial_x^2 B_0)^2 + 2(\partial_{xy}^2 B_0)^2 + (\partial_y^2 B_0)^2\, d\mathbf{r}.
\label{eq:bending}
\end{equation}
We keep $\lambda_{\mathrm{B0}}$ fixed during optimization. Too small a value fails to suppress the ambiguity and $B_0$ can drift, while too large a value over-smooths the final estimate.

\noindent\textbf{Learnable gradient isocenter.}
The rotated coordinate grid for orientation $n$ at phase-encoding angle $\theta_n$ is $\mathbf{r}'_n = \mathbf{R}(\theta_n)\,(\mathbf{r} - \mathbf{c}) + \mathbf{c}$, a rotation about an isocenter $\mathbf{c} = (c_x, c_y)$. The rotated gradients pivot about the gradient isocenter, which scanner hardware imperfections shift from the image matrix center by a few pixels, an amount not known a priori. A single orientation is invariant to this offset, but combining multiple rotated grids is not. A few-pixel isocenter error misaligns the grids and blurs the joint reconstruction. For multi-orientation acquisitions we therefore treat $\mathbf{c}$ as a learnable parameter, jointly optimized with the primitives, $B_0$, and $\Phi$.

\noindent\textbf{Partial-DFT EPI forward.}
The encoding $\mathcal{E}_{n,s}(B_0)$ of~\eqref{eq:encoding} is evaluated only at the acquired PE lines. Because its $B_0$ phase $e^{-i\,2\pi B_0\,t_{p,s}}$ varies from PE line to PE line (through $t_{p,s}$), the PE-direction transform admits no fast FFT; we compute it as a partial DFT, reducing the PE-direction cost from a full $N_y \times N_y$ DFT to a $P \times N_y$ product with $P \ll N_y$ at high acceleration.

\noindent\textbf{Inference.}
After convergence, the corrected image and $B_0$ map are obtained by querying the trained primitives on a Cartesian grid; being continuous, the primitives can also be sampled on grids finer than the acquisition matrix.

\section{Experimental Setup}
\label{sec:experiments}

\begin{figure}[!tbp]
    \centering
    \includegraphics[width=\linewidth]{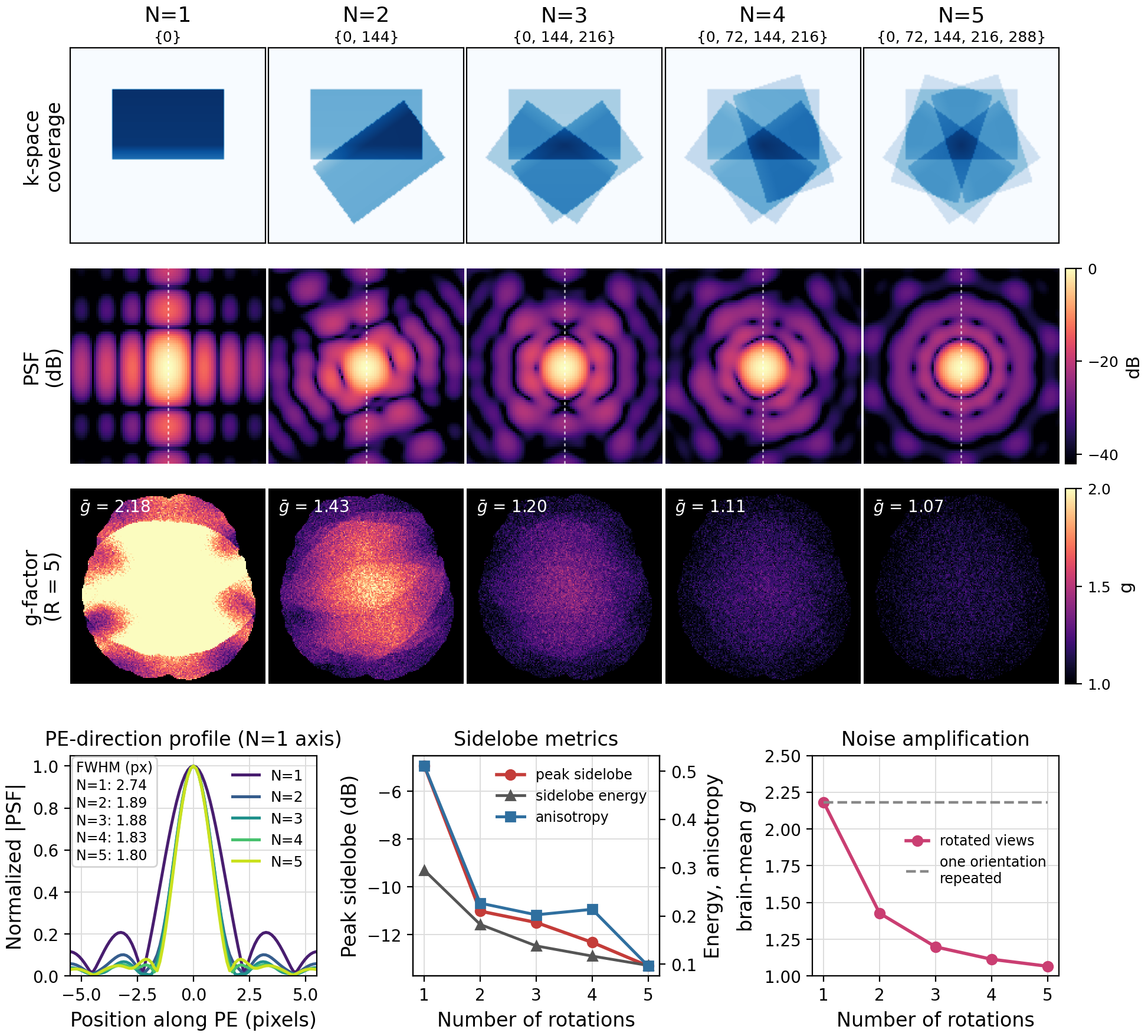}
    \caption{Acquisition point spread function (PSF) for $N_{\mathrm{rot}}=1$--$5$ rotated EPI views, with $5/8$ partial Fourier and a $T_2/T_2^\ast$ decay envelope along the phase-encoding direction ($T_2=70$\,ms, $T_2^\ast=36$\,ms, $T_\mathrm{RO}=70$\,ms; representative envelope parameters, not the literal echo-train length). Row 1: accumulated \textit{k}-space coverage. Row 2: 2D PSF in dB. Row 3: g-factor of blip-up/blip-down pairs at the same orientations ($R{=}5$, 32-channel coil, $b{=}0$, unregularized SENSE least squares without partial Fourier, 192 pseudo-replicas), relative to a fully sampled acquisition of equal duration and evaluated within the \textit{k}-space disk covered by every orientation, so that all $N_{\mathrm{rot}}$ are compared at the same isotropic resolution; $\bar{g}$ is the brain mean. Bottom row: PSF cross-section along the $N_{\mathrm{rot}}{=}1$ phase-encoding axis, with main-lobe FWHM (px) per $N_{\mathrm{rot}}$, three sidelobe metrics (peak sidelobe, integrated sidelobe energy, near-mainlobe angular CoV) versus $N_{\mathrm{rot}}$, and $\bar{g}$ versus $N_{\mathrm{rot}}$.}
    \label{fig:psf_analysis}
\end{figure}

\begin{figure*}[!tbp]
    \centering
    \includegraphics[width=\linewidth]{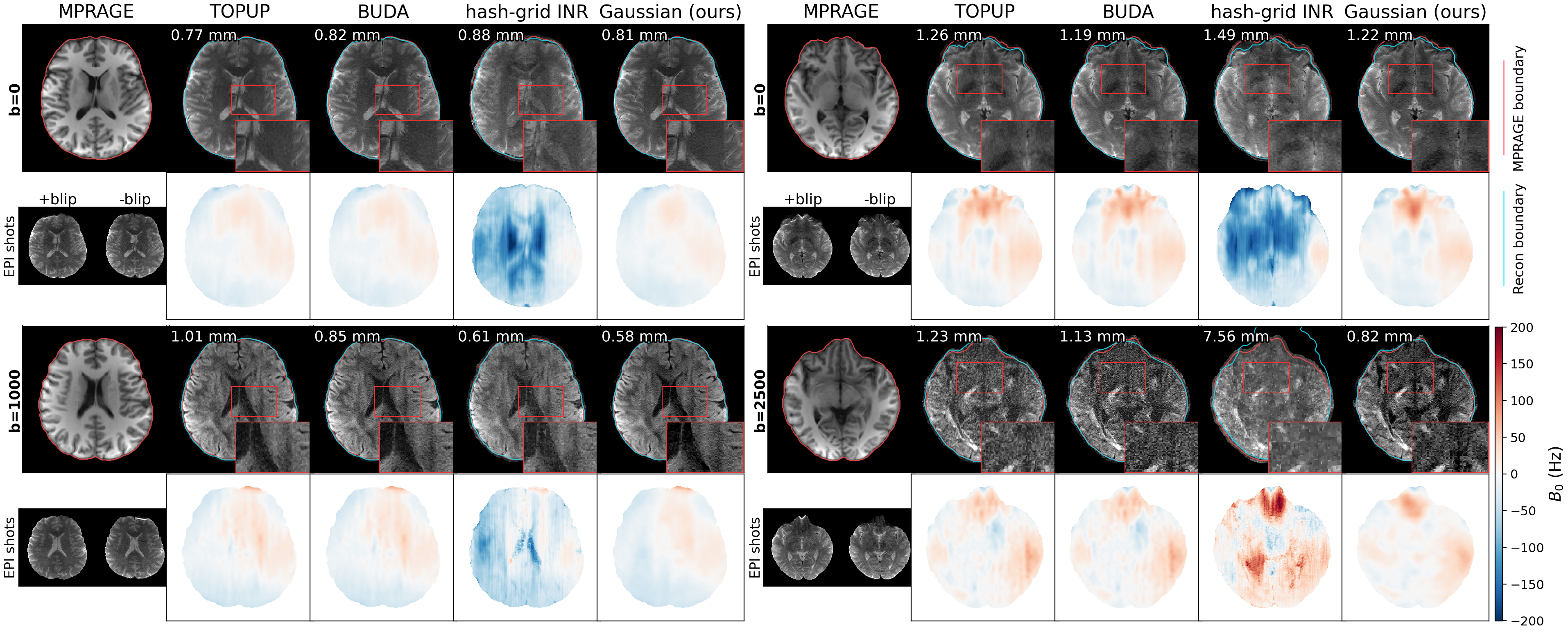}
    \caption{Single-orientation (blip-up/blip-down) correction on \textit{in vivo} brain diffusion EPI at $R{=}4$, a $2{\times}2$ grid: top row two $b{=}0$ slices, bottom row $b{=}1000$ and $b{=}2500$\,s/mm$^2$. Per case, the top sub-row is the corrected magnitude and the bottom the estimated $B_0$; columns are the MPRAGE reference ($\pm$blip CG-SENSE shots below), TOPUP, BUDA, hash-grid INR, and the proposed method. On magnitude panels, red marks the brain boundary of the MPRAGE and cyan the reconstruction's own, extracted in the same way on both sides; their gap is the residual distortion, and the number on each panel is the per-slice average symmetric surface distance in mm. All cases share the $B_0$ color scale.}
    \label{fig:comparison_1rot}
\end{figure*}

\subsection{Data and Implementation}
\label{sec:data_impl}

\paragraph{Data set 1}
All acceleration in this work is prospective. We acquired \textit{in vivo} brain diffusion EPI from five subjects on a 3\,T GE SIGNA UHP scanner (GE Healthcare, Waukesha, WI, USA) with a 32-channel head coil (Nova Medical). All scanning was performed under a protocol approved by the Stanford University Institutional Review Board, with written informed consent obtained from every participant. For each subject, all five rotated views ($0^\circ$, $72^\circ$, $144^\circ$, $216^\circ$, and $288^\circ$) were acquired together in one prospectively undersampled series at $b{=}0$, $1000$, and $2500$\,s/mm$^2$ (six diffusion-encoding directions at each non-zero $b$-value), with the phase-encoding direction rotating across successive TRs. Sequence parameters were TE $=69$ and $63$\,ms and TR $=2.83$ and $2.92$\,s at $R{=}4$ and $5$, respectively, echo spacing $1.296$\,ms, $220 \times 220$ matrix, FOV $220\times220$\,mm$^2$ ($1.0\times1.0$\,mm$^2$ in-plane), slice thickness $5$\,mm ($16$ slices), and $5/8$ partial Fourier. We acquired a low-resolution GRE auto-calibration (ACS) pre-scan, estimated the coil sensitivities from it with ESPIRiT~\cite{uecker2014espirit}, and reused the same calibration \textit{k}-space to initialize the image-primitive positions and weights. A single 3D MPRAGE~\cite{mugler1990three} was acquired as a distortion-free anatomical reference. Each orientation provides a blip-up and a blip-down single-shot readout (two shots). Each experiment then uses $N_{\mathrm{rot}}$ of the five acquired orientations, so paired-polarity reconstruction takes $2N_{\mathrm{rot}}$ shots and blip-up-only takes $N_{\mathrm{rot}}$. The single-orientation TOPUP and BUDA baselines use the $0^\circ$ blip-up/blip-down pair, the conventional un-rotated acquisition. Acceleration factors $R{=}4$ and $R{=}5$ were evaluated.

\paragraph{Rotated-view implementation}
Each view rotates the readout and phase-encoding gradient axes together in the logical gradient frame about the slice-select axis, which stays fixed. Only the Nyquist-ghost reference has to be repeated for each orientation; it is a single $2$\,s calibration TR, so five orientations add $8$\,s. The coil sensitivities need one low-resolution GRE auto-calibration scan, as in conventional EPI: the receive coils do not rotate with the views. A multi-orientation acquisition has the same number of TRs as multi-shot phase-segmented EPI with the same number of shots. For the 13 volumes of Data set 1 a series lasts $1.2$\,min for the single-orientation blip-up/blip-down pair and $6.1$\,min for five paired orientations at $R{=}4$ ($1.3$ and $6.3$\,min at $R{=}5$). The sequence ran within a maximum gradient amplitude of $40$\,mT/m and a slew rate of $100$\,T/m/s, within the specifications of widely deployed clinical systems such as the GE MR750 and Premier.

\paragraph{Data set 2}
An additional scan with the resolution and whole-brain coverage of a routine clinical diffusion protocol was acquired on the same scanner with 30 diffusion directions: one $b{=}0$ volume and 30 directions at $b{=}1000$\,s/mm$^2$, five rotated orientations each with a blip-up and a blip-down readout, prospective in-plane acceleration $R{=}3$, $56$ slices of $2$\,mm thickness, FOV $220\times220$\,mm$^2$ with a $132\times128$ (PE\,$\times$\,RO) matrix, TE $=62$\,ms, TR $=6.36$\,s, and echo spacing $0.976$\,ms. The readout was zero-padded from 128 to 132 samples, giving a square $1.67\times1.67$\,mm$^2$ in-plane grid. The 31 volumes take $32.9$\,min with five paired orientations and $6.6$\,min with the single-orientation pair.

\paragraph{Model}
The image is represented by $N_{\mathrm{img}} = 50{,}000$ Gaussian primitives and the $B_0$ field by $N_{\mathrm{B0}} = 1{,}000$. Each shot owns a complex-weighted Gaussian basis of $N_\varphi = 300$ primitives for its phase factor. The number of estimated parameters is $310{,}200$ for a single blip-up/blip-down pair and $327{,}002$ for five paired orientations.

\paragraph{Optimization}
We use the Adam optimizer~\cite{kingma2015adam} for 4,000 iterations; after each step the image weights are projected to be non-negative. The $B_0$ bending-energy coefficient is set at $\lambda_{\mathrm{B0}} \approx 1\!\times\!10^{-2}$ and, as in TOPUP, scaled by the current data-consistency residual at each iteration. The gradient isocenter is initialized by a Lucas-Kanade~\cite{lucas1981iterative} alignment of the single-orientation reconstructions and, for multi-orientation acquisitions, refined jointly during optimization.

\paragraph{Hardware and implementation}
All experiments run in PyTorch on an NVIDIA RTX A6000 GPU. Two custom CUDA kernels address the two bottlenecks.\footnote{Code, including the CUDA kernels, will be released at \url{https://github.com/wenqihuang/gaussian-epi-distortion}.} The first is primitive evaluation: rendering $N_{\mathrm{img}}=50{,}000$ anisotropic Gaussians on a $220\times220$ grid as a dense sum materializes a per-primitive, per-pixel intermediate of $2.4\times10^{9}$ entries that does not fit in GPU memory. The tile-based evaluator culls primitives whose footprint does not overlap a tile, so the cost scales with the footprint of each primitive rather than with the image size. The second is the EPI forward operator: the encoding is the partial DFT of Section~\ref{sec:optimization}, and a direct autograd implementation stores the per-line phase tensor for every coil and shot. The kernel pair fuses the off-resonance modulation with the line-selective DFT and implements the analytic adjoint as a matched backward pass, which removes these intermediates and runs the forward and backward passes $1.45\times$ faster than the PyTorch implementation. End-to-end, a typical $R{=}4$, $b{=}0$, six-shot ($N_{\mathrm{rot}}{=}3$) reconstruction of a 32-coil $220\times220$ slice converges in approximately 36\,s. Slices and diffusion directions are reconstructed independently; on one RTX A6000 running three to four jobs concurrently, the $56$ slices and $31$ volumes of Data set 2 took about $7$\,h at $N_{\mathrm{rot}}{=}1$ and $22$\,h at $N_{\mathrm{rot}}{=}5$.

\subsection{Baselines}
\label{sec:baselines}

We compare against three baselines that cover the main approaches to EPI distortion correction. \textbf{TOPUP}~\cite{andersson2003correct,smith2004advances} is the standard sequential method in FSL. Distorted EPI images are first reconstructed via CG-SENSE~\cite{pruessmann2001advances}, and TOPUP then estimates the off-resonance field from the blip-up/blip-down image pair (with the per-voxel displacement derived from it) and corrects in the image domain; it operates on a common Cartesian grid and cannot process the rotated acquisition grids of rotated-view EPI. \textbf{BUDA}~\cite{liao2019highly,liao2021distortion} combines blip-up and blip-down \textit{k}-space data in a single Hankel-structured low-rank reconstruction, with the $B_0$ field estimated separately via TOPUP and then held fixed during the reconstruction. \textbf{Hash-grid INR}~\cite{huang2025pinr} is a prior INR-based joint reconstruction that uses hash-grid encoded MLPs~\cite{muller2022instant} as the image and $B_0$ representations. The original work was demonstrated on simulated data and did not model per-shot phase variation; we adapted it to \textit{in vivo} data by adding the same per-shot phase factor used in the proposed method. The comparison isolates the effect of switching from an implicit (hash-grid MLP) to an explicit (Gaussian primitive) spatial representation within an otherwise matched joint-optimization framework. On Data set 2 all three baselines were run with the same settings, except that the TOPUP warp resolution and smoothing, which are specified in millimeters, were scaled to the $1.67$\,mm voxel.

\subsection{Evaluation}
\label{sec:evaluation}

The joint reconstruction is assessed on two fronts: the geometric accuracy of the distortion correction and the detail and noise of the reconstructed image. We quantify the geometry against an independent, distortion-free structural reference, as is standard in the EPI distortion-correction literature~\cite{andersson2003correct,schallmo2021assessing}; pixel-wise intensity measures (PSNR, SSIM) do not apply, since they would require a same-contrast distortion-free image that EPI cannot provide. The detail and noise, which lack any such reference, are instead compared qualitatively across methods. The structural reference is a single 3D MPRAGE, rigidly registered to the same-session GRE calibration scan, which is free of EPI distortion, with SynthMorph~\cite{hoffmann2021synthmorph} and averaged over the slice thickness. The corrected EPI and the MPRAGE ($T_1$) differ in contrast, so their intensities do not correspond; the brain outline, however, is a geometric feature common to both, and EPI distortion displaces it along the phase-encoding direction, so the agreement between the two outlines measures the residual distortion. We skull-strip each method's corrected volume and the registered MPRAGE with the same tool, so that the extraction bias is common-mode. For the $b{=}0$ comparison, SynthStrip~\cite{hoopes2022synthstrip} generates masks that include the extracerebral cerebrospinal fluid (CSF) on both the MPRAGE and the corrected EPI; for the diffusion-weighted images ($b>0$), HD-BET~\cite{isensee2019hdbet} delineates the brain parenchyma on both, excluding the extracerebral CSF. We compare their brain boundaries in-plane, per slice: Dice of the filled masks, average symmetric surface distance (ASSD), and $95$th-percentile Hausdorff distance (HD95). We also report wall-clock reconstruction time per slice (bottom row of Table~\ref{tab:main}).

On Data set 2, diffusion tensors were fit with MRtrix3~\cite{tournier2019mrtrix} (weighted least squares) after Gibbs-ringing removal~\cite{kellner2016gibbs}, inside one brain mask common to all methods: the subject's MPRAGE, segmented with SynthSeg~\cite{billot2023synthseg}, intersected with the EPI signal support of every method. White matter is the SynthSeg cerebral white-matter label eroded by one voxel. We report white-matter mean FA and MD and two reference-free precision measures from tensors fit to two interleaved 15-direction halves of the data, $A$ and $B$: $\sigma_{\mathrm{FA}}$, the white-matter median of $|\mathrm{FA}_A-\mathrm{FA}_B|/\sqrt{2}$, and the white-matter median angle between the principal eigenvectors of $A$ and $B$. Fiber orientation distributions were estimated with single-shell constrained spherical deconvolution (CSD)~\cite{tournier2007robust} (maximum harmonic order 6), and their precision with a wild bootstrap (1000 resamples, slices 18 and 28)~\cite{whitcher2008wild,jeurissen2011probabilistic}, reported as the $95\%$ cone of uncertainty (U95) of the first fiber direction~\cite{jones2003determining} in the eroded white matter.

\begin{figure*}[!tbp]
    \centering
    \includegraphics[width=\linewidth]{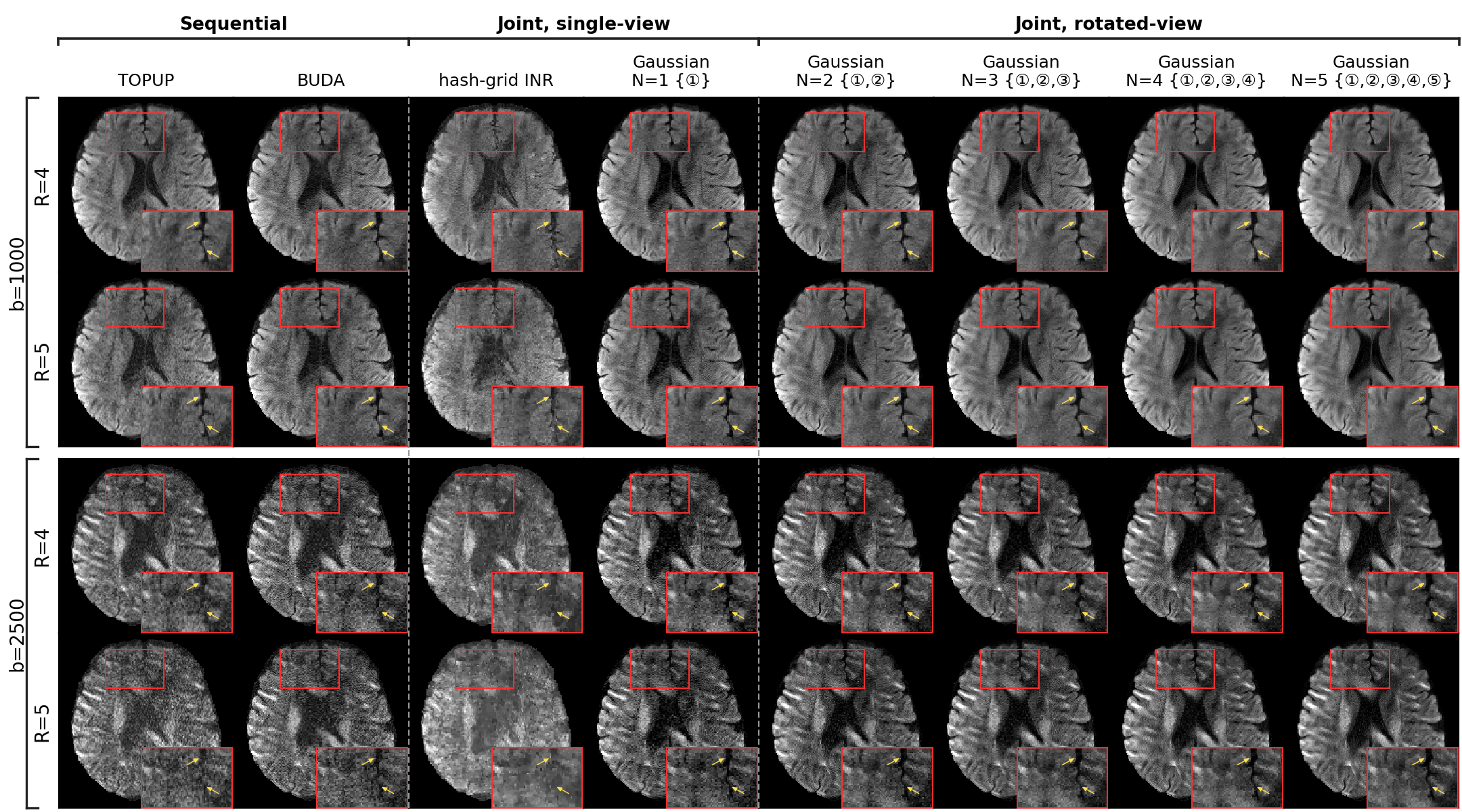}
    \caption{Reconstructions at in-plane acceleration $R\in\{4,5\}$ for $b{=}1000$ and $b{=}2500$\,s/mm$^2$. Columns are grouped as sequential baselines (TOPUP, BUDA), joint single-view (hash-grid INR, Gaussian $N_{\mathrm{rot}}{=}1$), and joint rotated-view (Gaussian $N_{\mathrm{rot}}{\ge}2$). Insets enlarge the boxed frontal region; yellow arrows mark the interhemispheric fissure.}
    \label{fig:acceleration}
\end{figure*}

\section{Results}
\label{sec:results}

\subsection{Single-Orientation Comparison}
\label{sec:results_1rot}

We first evaluate the proposed method under the standard blip-up/blip-down protocol with a single phase-encoding orientation. All methods consume the same two shots, so differences come from the estimator and the representation alone.

Fig.~\ref{fig:comparison_1rot} shows representative slices at $R{=}4$ across $b$. At low $b$ the parallel-imaging input is clean and all methods correct the distortion comparably on the brain-boundary metric (Dice $\approx0.99$; Table~\ref{tab:main}), but they differ in fine detail and diverge as $b$ rises. TOPUP is the smoothest, because it estimates the $B_0$ field from the blip-up/blip-down CG-SENSE reconstructions and applies it as a fixed input, so input imperfections bias the field and propagate into the image, which is then warped by image-domain B-spline interpolation, a low-pass step that smooths fine detail. BUDA recovers detail comparable to the proposed method, since both reconstruct directly in \textit{k}-space, but it inherits the same smooth $B_0$ from TOPUP, which limits its distortion correction and leaves its boundary increasingly behind the proposed method as $b$ rises. The hash-grid INR shows visible artifacts in most cases. The same INR, effective on synthetic data in the conference version (which had no per-shot phase), overfits once adapted \textit{in vivo}, where the image, $B_0$, and added per-shot phase are all high-capacity MLPs and the $B_0$ and phase leak into each other. The proposed method matches BUDA's detail while keeping the most spatially coherent $B_0$ and, at $b>0$, the closest boundary alignment; at $b{=}0$, where the input is clean, TOPUP, BUDA, and the proposed method lie within $0.1$\,mm of each other.

\subsection{Acceleration Factor}
\label{sec:results_acceleration}

Fig.~\ref{fig:acceleration} compares $R{=}4$ and $R{=}5$ at $b\in\{1000, 2500\}$\,s/mm$^2$. The sequential baselines become noisier with $R$, because the g-factor noise amplification (and any residual aliasing) of the CG-SENSE input grows with acceleration. The noise is most visible at $b{=}2500$, where SNR is lowest, while their boundary agreement shows no consistent trend between $R{=}4$ and $5$ (Table~\ref{tab:main}).

By contrast, the proposed method stays stable across $R$ and $b$ on the brain-boundary metric (Dice $0.985$--$0.991$ for every $N_{\mathrm{rot}}$, Table~\ref{tab:main}); the single-orientation ($N_{\mathrm{rot}}{=}1$) reconstruction alone matches the baselines at $b{=}0$ and exceeds them at $b>0$. Added orientations change Dice little but tighten the boundary surface distances, most at $R{=}5$, $b{=}2500$ ($N_{\mathrm{rot}}{=}1\!\to\!3$: HD95 $2.49\!\to\!2.02$\,mm) and in the high-susceptibility regions (Section~\ref{sec:results_regions}), with $N_{\mathrm{rot}}{=}3$ and $5$ nearly identical (Table~\ref{tab:main}); their larger benefit is in intra-brain detail and PSF isotropy (Fig.~\ref{fig:psf_analysis}), which this boundary metric does not probe. The advantage over the sequential pipeline vanishes at $b{=}0$ and grows with $b$-value; at $b{=}2500$ the proposed method attains the best agreement on every metric at both accelerations (ASSD $0.72$--$0.76$ versus $0.95$ for BUDA and $1.15$\,mm for TOPUP at $R{=}4$; $0.86$--$0.90$ versus $1.16$ and $1.00$\,mm at $R{=}5$). The hash-grid INR falls off most sharply at $b{=}2500$ (Dice $0.93$ and $0.92$ at $R{=}4$ and $5$).

\subsection{Multi-Orientation: Image Quality and Single-Polarity Viability}
\label{sec:results_multirot}

\begin{figure*}[!tbp]
    \centering
    \includegraphics[width=\linewidth]{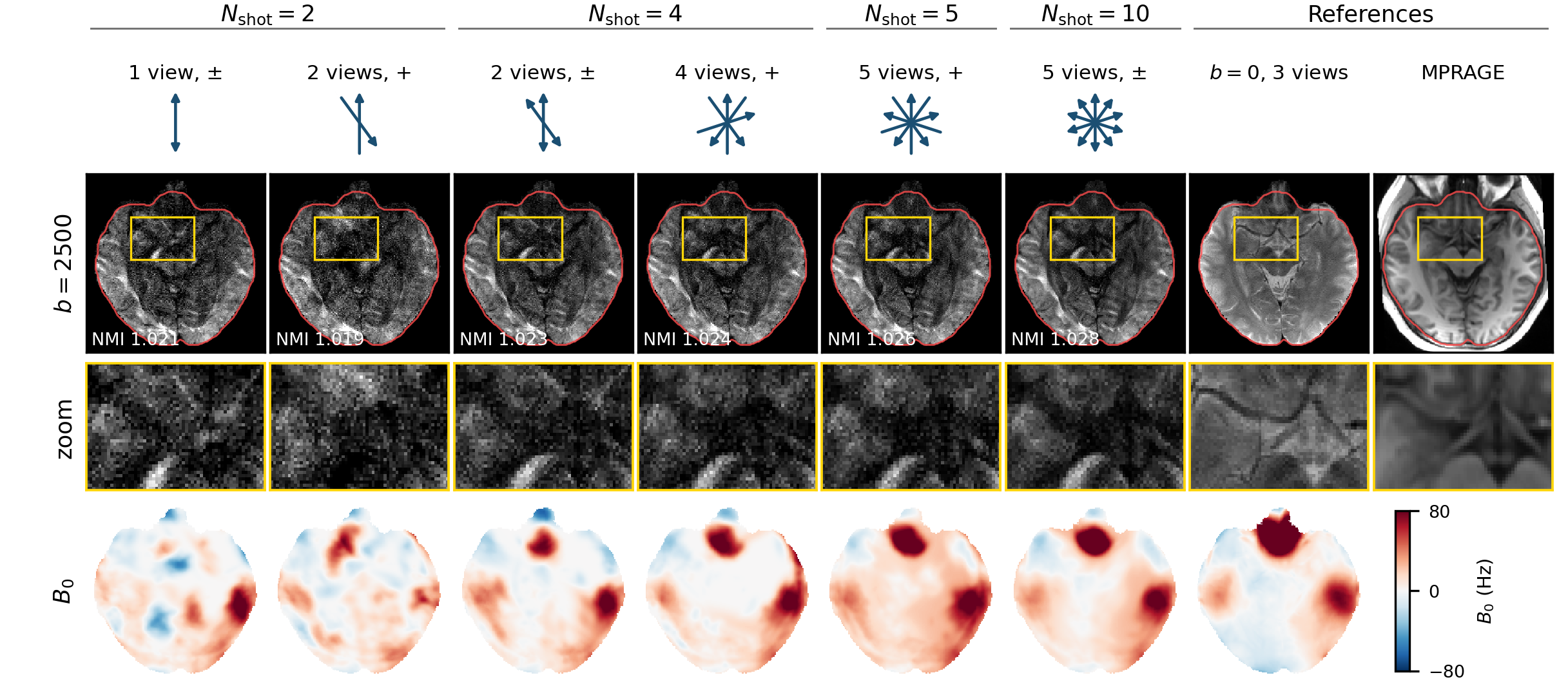}
    \caption{Acquisitions at matched shot count $N_{\mathrm{shot}}$ ($R{=}5$, $b{=}2500$\,s/mm$^2$, slice 2 near the frontal sinus), all reconstructed with the proposed method. Columns are grouped by $N_{\mathrm{shot}}$; the icons show the phase-encoding orientations, with double arrows for blip-up/blip-down pairs ($\pm$) and single arrows for blip-up only ($+$). Rows: the $b{=}2500$ image with the brain outline of the registered MPRAGE (red) and the NMI with the GRE calibration image, the boxed region enlarged, and the estimated $B_0$. References: the $b{=}0$, three-view, $R{=}4$ reconstruction of the same slice with its $B_0$, and the registered MPRAGE. All arms share one magnitude window and one $B_0$ color scale.}
    \label{fig:abl_matched}
\end{figure*}

\begin{table}[t]
\centering
\caption{Brain-boundary agreement with the registered MPRAGE reference: per-slice Dice/ASSD/HD95 from the brain masks of Section~\ref{sec:evaluation}, mean$\pm$std over slices; best per column in bold. $b$ in s/mm$^2$; INR = hash-grid INR; $N_{\mathrm{rot}}$ = number of jointly reconstructed EPI orientations (proposed method). Times are per-slice wall-clock (RTX A6000; FSL TOPUP steps on the CPU), $R$/$b$-independent, for the full \textit{k}-space-to-image pipeline: TOPUP/BUDA include CG-SENSE reconstruction (BUDA$^{\ast}$ also its internal TOPUP field estimation), as does the proposed direct method. Ours ($N_{\mathrm{rot}}{=}1$) uses a single orientation with blip-up/blip-down polarity, the same two shots that TOPUP and BUDA use.}
\label{tab:main}
\resizebox{\linewidth}{!}{%
\begin{tabular}{c l cccccc}
\toprule
& \multirow{2}{*}{Method} & \multicolumn{3}{c}{$R{=}4$} & \multicolumn{3}{c}{$R{=}5$} \\
\cmidrule(lr){3-5}\cmidrule(lr){6-8}
& & $b{=}0$ & $b{=}1000$ & $b{=}2500$ & $b{=}0$ & $b{=}1000$ & $b{=}2500$ \\
\midrule
\multirow{6}{*}{\rotatebox[origin=c]{90}{Dice\,$\uparrow$}}
  & TOPUP           & \textbf{0.987}$\pm$0.004 & 0.987$\pm$0.001 & 0.984$\pm$0.001 & 0.986$\pm$0.005 & 0.990$\pm$0.002 & 0.986$\pm$0.003 \\
& BUDA            & \textbf{0.987}$\pm$0.004 & 0.988$\pm$0.001 & 0.987$\pm$0.007 & 0.985$\pm$0.005 & 0.990$\pm$0.002 & 0.984$\pm$0.007 \\
& INR             & 0.983$\pm$0.004 & 0.976$\pm$0.018 & 0.929$\pm$0.019 & 0.978$\pm$0.004 & 0.986$\pm$0.003 & 0.915$\pm$0.027 \\
& Ours ($N_{\mathrm{rot}}{=}1$) & 0.986$\pm$0.004 & \textbf{0.991}$\pm$0.002 & \textbf{0.990}$\pm$0.002 & 0.985$\pm$0.004 & \textbf{0.991}$\pm$0.002 & \textbf{0.988}$\pm$0.005 \\
& Ours ($N_{\mathrm{rot}}{=}3$) & \textbf{0.987}$\pm$0.004 & \textbf{0.991}$\pm$0.002 & \textbf{0.990}$\pm$0.003 & \textbf{0.987}$\pm$0.004 & 0.990$\pm$0.003 & \textbf{0.988}$\pm$0.005 \\
& Ours ($N_{\mathrm{rot}}{=}5$) & \textbf{0.987}$\pm$0.004 & \textbf{0.991}$\pm$0.003 & \textbf{0.990}$\pm$0.003 & 0.986$\pm$0.005 & 0.989$\pm$0.004 & \textbf{0.988}$\pm$0.005 \\
\midrule
\multirow{6}{*}{\rotatebox[origin=c]{90}{ASSD (mm)\,$\downarrow$}}
  & TOPUP           & 0.97$\pm$0.26 & 0.94$\pm$0.07 & 1.15$\pm$0.10 & 1.05$\pm$0.29 & 0.74$\pm$0.12 & 1.00$\pm$0.18 \\
& BUDA            & 0.96$\pm$0.24 & 0.86$\pm$0.08 & 0.95$\pm$0.51 & 1.09$\pm$0.28 & 0.72$\pm$0.10 & 1.16$\pm$0.49 \\
& INR             & 1.25$\pm$0.23 & 1.73$\pm$1.23 & 5.33$\pm$1.68 & 1.61$\pm$0.21 & 1.01$\pm$0.19 & 6.29$\pm$2.17 \\
& Ours ($N_{\mathrm{rot}}{=}1$) & 1.05$\pm$0.26 & \textbf{0.65}$\pm$0.10 & 0.76$\pm$0.12 & 1.09$\pm$0.26 & \textbf{0.70}$\pm$0.13 & 0.87$\pm$0.34 \\
& Ours ($N_{\mathrm{rot}}{=}3$) & \textbf{0.94}$\pm$0.23 & 0.66$\pm$0.12 & \textbf{0.72}$\pm$0.15 & \textbf{0.98}$\pm$0.24 & 0.74$\pm$0.18 & \textbf{0.86}$\pm$0.27 \\
& Ours ($N_{\mathrm{rot}}{=}5$) & 0.96$\pm$0.25 & 0.66$\pm$0.14 & 0.73$\pm$0.17 & 0.99$\pm$0.27 & 0.79$\pm$0.24 & 0.90$\pm$0.30 \\
\midrule
\multirow{6}{*}{\rotatebox[origin=c]{90}{HD95 (mm)\,$\downarrow$}}
  & TOPUP           & 2.86$\pm$1.35 & 2.12$\pm$0.26 & 2.46$\pm$0.32 & 2.89$\pm$1.06 & 1.87$\pm$0.55 & 2.64$\pm$0.60 \\
& BUDA            & 2.77$\pm$1.23 & 2.09$\pm$0.34 & 3.27$\pm$3.88 & 2.86$\pm$1.05 & 1.84$\pm$0.45 & 3.86$\pm$2.32 \\
& INR             & 3.99$\pm$1.53 & 5.18$\pm$3.95 & 18.96$\pm$6.38 & 4.29$\pm$1.29 & 2.50$\pm$0.52 & 16.50$\pm$4.52 \\
& Ours ($N_{\mathrm{rot}}{=}1$) & 2.91$\pm$1.31 & \textbf{1.59}$\pm$0.53 & 1.87$\pm$0.39 & 2.77$\pm$0.97 & \textbf{1.50}$\pm$0.42 & 2.49$\pm$1.86 \\
& Ours ($N_{\mathrm{rot}}{=}3$) & 2.70$\pm$1.08 & \textbf{1.59}$\pm$0.46 & \textbf{1.82}$\pm$0.45 & \textbf{2.54}$\pm$0.92 & 1.58$\pm$0.41 & \textbf{2.02}$\pm$0.63 \\
& Ours ($N_{\mathrm{rot}}{=}5$) & \textbf{2.67}$\pm$1.23 & 1.62$\pm$0.45 & 1.85$\pm$0.50 & 2.61$\pm$1.06 & 1.73$\pm$0.52 & 2.28$\pm$0.80 \\
\midrule
\multicolumn{2}{l}{Time per slice (s)\,$\downarrow$} & \multicolumn{6}{c}{TOPUP 137 \quad BUDA 140$^{\ast}$ \quad INR 68 \quad Ours $N_{\mathrm{rot}}{=}1/3/5$: \textbf{24}/36/59} \\
\bottomrule
\end{tabular}%
}
\end{table}

Fig.~\ref{fig:abl_matched} compares acquisitions at equal shot count $N_{\mathrm{shot}}$ at the most demanding setting, $R{=}5$ and $b{=}2500$\,s/mm$^2$. Within $N_{\mathrm{shot}}{=}2$ and $4$, one or two paired blip-up/blip-down views are set against twice as many blip-up-only rotated views; $N_{\mathrm{shot}}{=}5$ and $10$ use all five orientations, blip-up only or paired. Noise falls as the shots are spread over more orientations, and five views give the cleanest image. At equal shot count the blip-up-only rotated views reach the image quality of the paired views. The field maps show where the acquisitions differ: the frontal susceptibility hotspot of the $b{=}0$ reference is almost absent with one paired view and is recovered from two paired views or four blip-up-only views onward ($N_{\mathrm{shot}}{=}4$).

The acquisition PSF explains this geometrically (Fig.~\ref{fig:psf_analysis}). At $N_{\mathrm{rot}}=1$, partial-Fourier ringing and $T_2^\ast$ broadening both concentrate along the phase-encoding axis, producing a strongly anisotropic PSF with a heavy sidelobe ridge in one direction. Additional orientations redistribute these effects, narrowing the main-lobe FWHM along the $N_{\mathrm{rot}}{=}1$ PE axis from $2.74$\,px to $1.80$\,px at $N_{\mathrm{rot}}{=}5$ (largely converged by $N_{\mathrm{rot}}{=}3$) and lowering the peak sidelobe and integrated sidelobe energy monotonically.

\subsection{Matched Acquisition Time}
\label{sec:results_ablation}

To separate the effect of the rotated views from that of the extra data, we compare the acquisitions of Fig.~\ref{fig:abl_matched} at matched shot count on all 16 slices ($R{=}5$, $b{=}2500$\,s/mm$^2$). Field deviations are in-brain root-mean-square differences from the field estimated from the $b{=}0$, three-view, $R{=}4$ acquisition of the same slice, reported as mean$\pm$std over slices.

With more shots spent on distinct orientations the field approaches the reference: the deviation falls from $16.2\pm8.4$\,Hz at $N_{\mathrm{shot}}{=}2$ (one paired view) to $14.3\pm5.7$\,Hz at $N_{\mathrm{shot}}{=}5$ (five blip-up-only views) and $10.9\pm3.8$\,Hz at $N_{\mathrm{shot}}{=}10$ (five paired views). At equal shot count, rotated blip-up-only views and paired views perform alike: $17.1\pm7.6$ versus $16.2\pm8.4$\,Hz at $N_{\mathrm{shot}}{=}2$, and $13.2\pm5.5$ versus $14.1\pm7.0$\,Hz at $N_{\mathrm{shot}}{=}4$. The NMI with the GRE calibration image rises with the shot count, from $1.019$--$1.021$ at $N_{\mathrm{shot}}{=}2$ to $1.023$--$1.024$ at $N_{\mathrm{shot}}{=}4$, $1.026$ at $N_{\mathrm{shot}}{=}5$, and $1.028$ at $N_{\mathrm{shot}}{=}10$; within a shot count the differences lie in the third decimal.

\subsection{Region-Stratified Boundary Agreement}
\label{sec:results_regions}
Whole-brain averages can mask errors confined to the regions where distortion is largest. We therefore split the boundary distances by region with the SynthSeg parcellation~\cite{billot2023synthseg} of the registered MPRAGE: the high-susceptibility (HS) region comprises the orbitofrontal cortex and the inferior and medial temporal lobe (inferior temporal, fusiform, entorhinal, and parahippocampal gyri), dilated in-plane by $5$\,mm, and the rest of the brain boundary forms the remainder. Boundary distances in both directions are pooled over slices, and their mean (ASSD) is reported per region.

At $R{=}4$ and $b{=}0$, the HS ASSD of the single-orientation methods is about twice that of the remainder ($1.96$--$2.05$\,mm for TOPUP, BUDA, and the proposed method with $N_{\mathrm{rot}}{=}1$, against $0.85$--$0.93$\,mm). Rotated views lower it to $1.78$ and $1.76$\,mm at $N_{\mathrm{rot}}{=}3$ and $5$ while the remainder stays at $0.85$--$0.86$\,mm, so the gain from the rotated views lies where the distortion is largest. At $b{=}2500$, the error of BUDA concentrates in the HS region ($2.68$\,mm, against $0.74$\,mm in the remainder), whereas TOPUP ($1.20$\,mm) and the proposed reconstructions ($0.90$--$1.03$\,mm) exceed their remainder values by at most $0.3$\,mm; at $b{=}1000$ the HS ASSD of TOPUP, BUDA, and the proposed reconstructions lies within $0.87$--$1.02$\,mm. The hash-grid INR exceeds $2.8$\,mm in the HS region in every $R{=}4$ cell. At $R{=}5$ the pattern is the same: at $b{=}0$ the rotated views lower the HS ASSD from $1.95$--$1.97$\,mm (TOPUP, BUDA, $N_{\mathrm{rot}}{=}1$) to $1.55$--$1.56$\,mm, and at $b{=}2500$ from $1.80$ (TOPUP), $3.80$ (BUDA), and $1.63$\,mm ($N_{\mathrm{rot}}{=}1$) to $0.96$--$0.99$\,mm.

\begin{table}[t]
\centering
\caption{Diffusion tensor metrics on Data set 2 (whole brain, $R{=}3$, $b{=}1000$\,s/mm$^2$, 30 directions). WM: mean over the eroded SynthSeg cerebral white matter. $\sigma$: direction-split precision, $|X_A-X_B|/\sqrt{2}$ between two interleaved 15-direction fits, WM median. MD in $10^{-3}$\,mm$^2$/s.}
\label{tab:dti}
\begin{tabular}{l cc cc}
\toprule
Method & FA$_{\mathrm{WM}}$ & MD$_{\mathrm{WM}}$ & $\sigma_{\mathrm{FA}}\,\downarrow$ & $\sigma_{\mathrm{MD}}\,\downarrow$ \\
\midrule
TOPUP & 0.499 & 0.729 & 0.0310 & 0.0161 \\
BUDA & 0.507 & 0.744 & 0.0342 & 0.0182 \\
INR ($N_{\mathrm{rot}}{=}1$) & 0.466 & 0.674 & 0.0505 & 0.0223 \\
Ours ($N_{\mathrm{rot}}{=}1$) & 0.502 & 0.756 & 0.0295 & 0.0156 \\
Ours ($N_{\mathrm{rot}}{=}3$) & 0.491 & 0.808 & 0.0325 & 0.0177 \\
Ours ($N_{\mathrm{rot}}{=}5$) & 0.491 & 0.800 & \textbf{0.0278} & \textbf{0.0147} \\
\bottomrule
\end{tabular}
\end{table}

\begin{figure*}[!tbp]
    \centering
    \includegraphics[width=\linewidth]{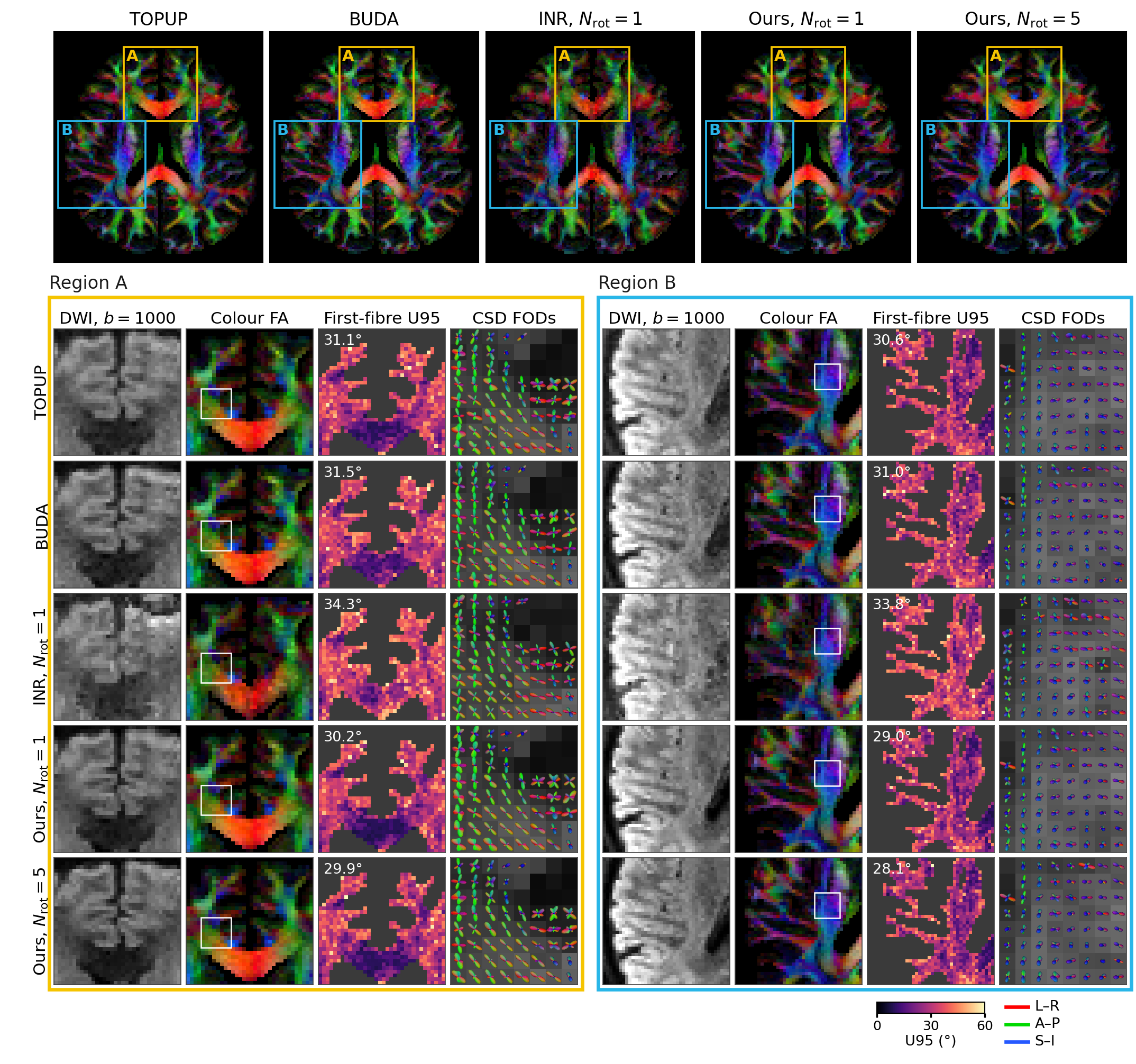}
    \caption{Data set 2 (whole brain, $R{=}3$, $1.67\times1.67\times2$\,mm$^3$, $b{=}1000$\,s/mm$^2$), slice 28. Top: color FA of each method with regions A (genu of the corpus callosum and frontal white matter) and B (corona radiata); color FA is shown inside the SynthSeg brain mask with the CSF excluded, the same mask for every method. Below, one row per method and, for each region, one diffusion-weighted image, the color FA, the bootstrap U95 of the first fiber direction in cerebral white matter with its median, and the CSD fiber orientation distributions in the $8\times8$-voxel window marked in the color FA (white matter only, colored by direction: red L--R, green A--P, blue S--I; viewed from superior).}
    \label{fig:diffusion}
\end{figure*}

\subsection{Whole-Brain Diffusion at $R{=}3$}
\label{sec:results_dti}

Data set 2 tests a routine operating point: moderate acceleration, $2$\,mm slices, whole-brain coverage, and $b{=}1000$\,s/mm$^2$ (Fig.~\ref{fig:diffusion}). Here the parallel-imaging input of the sequential pipeline is well conditioned, and the single-view joint reconstruction gains little over BUDA in the images; the hash-grid INR shows blotchy artifacts in the diffusion-weighted images and in the color FA.

The diffusion tensor metrics (Table~\ref{tab:dti}) and the fiber orientations (Fig.~\ref{fig:diffusion}) give the same order. The five-view reconstruction gives the most precise tensors of all methods, $\sigma_{\mathrm{FA}}=0.0278$, $10\%$ below TOPUP ($0.0310$) and $19\%$ below BUDA ($0.0342$), and the smallest split angle between principal eigenvectors ($6.9^\circ$ versus $8.0^\circ$ and $8.9^\circ$). The hash-grid INR, run with the same estimator, data, and per-shot phase as the proposed method at $N_{\mathrm{rot}}=1$, gives the least precise tensors ($\sigma_{\mathrm{FA}}=0.0505$, split angle $14.9^\circ$) and the lowest white-matter FA ($0.466$). The white-matter FA of the other methods lies within $0.016$ ($0.491$--$0.507$). White-matter MD of the proposed reconstructions ($0.76$--$0.81\times10^{-3}$\,mm$^2$/s) is $2$--$11\%$ above TOPUP and BUDA.

Fig.~\ref{fig:diffusion} shows the fiber orientations in two regions, the genu of the corpus callosum with the adjoining frontal white matter (A) and the corona radiata (B). TOPUP, BUDA, and the proposed reconstructions resolve the same single- and crossing-fiber configurations, whereas the hash-grid INR adds spurious secondary lobes. The bootstrap uncertainty of the first fiber direction is lowest with five views: the white-matter median U95 over the two bootstrapped slices is $26.1^\circ$, against $28.4^\circ$ for TOPUP, $29.7^\circ$ for BUDA, and $32.2^\circ$ for the INR.

\section{Discussion}
\label{sec:discussion}

\subsection{Findings and Interpretation}
\label{sec:disc_findings}
The proposed method's advantage over the sequential pipeline is largest where the parallel-imaging input is most stressed (high $b$; Table~\ref{tab:main}) and vanishes at $b{=}0$ where that input is already clean. The joint model fits all shots to one shared image, and because the blip-up, blip-down, and rotated readouts sample complementary \textit{k}-space, that image is constrained by their pooled coverage rather than by any single accelerated shot.

The $N_{\mathrm{rot}}{=}3$ practical operating point appears consistently across two independent analyses, PSF isotropy (Fig.~\ref{fig:psf_analysis}) and \textit{in vivo} image quality (Fig.~\ref{fig:acceleration}), suggesting three orientations balance geometric diversity against scan time.

\subsection{Explicit Primitives versus Implicit Neural Representations}
\label{sec:disc_inr}
INRs~\cite{sitzmann2020implicit,muller2022instant} are parameter-efficient when the target signal is approximately stationary, but for scan-specific joint reconstruction three properties of explicit primitives matter more. First, each primitive has compact spatial support, so a gradient update propagates only to primitives whose footprint overlaps the gradient pixel; this keeps the optimization local and avoids the hash-collision-mediated entanglement that can leave the hash-grid INR baseline in poorly resolved solutions in which the $B_0$ and per-shot phase leak into each other (Fig.~\ref{fig:comparison_1rot}). Second, overfitting is controlled by a single interpretable knob, the minimum scale floor (Section~\ref{sec:optimization}), which directly caps per-primitive frequency content; the analogous safeguard for an INR (weight decay, network width, hash table size) is less direct and harder to tune at deployment. Third, primitive parameters (position, anisotropic scale, orientation) are inspectable per primitive, whereas the same information is distributed implicitly across the weights of an INR.

\subsection{Practical Implications and Applicability}
\label{sec:disc_practical}
Decoupling distortion correction from the quality of an intermediate parallel-imaging reconstruction has two practical consequences. Higher in-plane acceleration stays usable, with the shorter echo trains and reduced distortion it brings. Angular diversity can also replace opposite-polarity pairing, so a single-polarity blip-up-only acquisition halves the diffusion-encoding shots. The same physics forward model over an explicit, continuous primitive field, fit directly to \textit{k}-space, should also apply to other $B_0$-distorted EPI such as functional and high-resolution structural imaging.

\noindent\textbf{Minimum acquisition and existing data.}
The minimum input is $N_{\mathrm{rot}}{=}1$ with $\mathcal{P}=\{+,-\}$, a single orientation with reversed polarity for every diffusion volume, as acquired, for example, in the Human Connectome Project~\cite{sotiropoulos2013advances}. In that configuration the rotated-view benefit is unavailable, but joint estimation still removes the dependence on an intermediate parallel-imaging reconstruction, and the $N_{\mathrm{rot}}{=}1$ rows of Table~\ref{tab:main} quantify this case. Beyond the reversed-polarity pair, the method needs coil sensitivities from any calibration region or ACS scan, and the sequence timing that enters the forward model (echo spacing, blip polarity, partial-Fourier coverage); it needs neither a measured field map nor training data. It does require the raw multi-coil \textit{k}-space, which most public diffusion data sets do not distribute; archives that store only reconstructed magnitude images cannot be reprocessed with it.

\noindent\textbf{Hardware.}
The reconstruction depends on the hardware only through the sequence timing and the gradient-isocenter offset between rotated views, which is estimated from the data.

\noindent\textbf{Slice thickness.}
The reconstruction is 2D and slice-independent, so thinner slices lower the SNR per slice without changing the forward model. Data set 2 uses $2$\,mm slices with whole-brain coverage (Section~\ref{sec:results_dti}). Thinner slices also reduce the variation of the field across the slice thickness, which displaces different depths of a slice by different amounts and which no method with one field value per voxel can correct.

\subsection{Limitations and Future Work}
\label{sec:disc_limits}
Completing the physics model is the most immediate extension: a 3D primitive representation with through-slice coupling and inter-slice motion correction. A harder, open question is how far acceleration and $b$-value can be pushed, since we tested only up to $R{=}5$ and $b{=}2500$\,s/mm$^2$, beyond which the $B_0$ and per-shot-phase ambiguity intensifies and better-designed priors are likely needed. For deployment, the iterative joint optimization keeps reconstruction off real-time at $\sim$24--59\,s per slice (one to five orientations), so accelerating it matters. Data set 2 is single-shell ($b{=}1000$\,s/mm$^2$, 30 directions); it supports DTI and single-shell CSD. Replacing the image representation inside TOPUP or BUDA was not tested, as it would still estimate the field from two separately reconstructed images, the step that joint estimation removes. No separately measured field map was acquired, so field accuracy is assessed against a high-SNR $b{=}0$ estimate from the same estimator rather than against an independent measurement.

\section{Conclusion}
\label{sec:conclusion}

We presented a physics-informed framework that jointly estimates the $B_0$ field and the distortion-free image directly from rotated multi-view EPI \textit{k}-space, removing the sequential pipeline's dependence on an intermediate parallel-imaging reconstruction. The image, the $B_0$ field, and the per-shot phase share one explicit, continuous Gaussian-primitive representation, with the image constrained to be real and non-negative and the phase carried by a unit-modulus factor; a bending-energy prior and the angular diversity of rotated views together reduce the single-orientation high-$b$ ambiguity. On \textit{in vivo} brain diffusion EPI the method attains the closest brain-boundary agreement with a distortion-free reference, with its largest gains at high $b$, where sequential pipelines are most fragile. On a whole-brain 30-direction acquisition at a routine acceleration, five-view reconstruction gives the most precise diffusion tensor estimates of the compared methods.

\bibliographystyle{IEEEtran}
\bibliography{references_r1}

\end{document}